\documentclass[runningheads]{llncs}
\usepackage[T1]{fontenc}
\usepackage{graphicx}
\usepackage{amsmath}
\usepackage{amssymb}
\usepackage{booktabs}
\usepackage{multirow}
\usepackage{caption}
\usepackage{subcaption}
\usepackage[hyphens]{url}
\usepackage{tabularx} 
\usepackage{array}
\newcolumntype{Y}{>{\centering\arraybackslash}X} 

\begin{document}
\title{Breast-Rehab: A Postoperative Breast Cancer Rehabilitation Training Assessment System Based on Human Action Recognition}
\titlerunning{Breast-Rehab}
%
\author{Zikang Chen\inst{1} \and
Tan Xie\inst{2} \and
Qinchuan Wang\inst{2} \and
Heming Zheng \inst{2} \and
Xudong Lu \inst{1}
}
\authorrunning{Zikang. et al.}
%
\institute{ Zhejiang University, Hangzhou 310058, China \and Zhejiang University School of Medicine Sir Run Run Shaw Hospital, Hangzhou 310058, China\\
\email{lvxd@zju.edu.cn}}
\maketitle              
\begin{abstract}
Postoperative upper limb dysfunction is prevalent among breast cancer survivors, yet their adherence to at-home rehabilitation exercises is low amidst limited nursing resources. The hardware overhead of commonly adopted VR-based mHealth solutions further hinders their widespread clinical application. Therefore, we developed Breast-Rehab, a novel, low-cost mHealth system to provide patients with out-of-hospital upper limb rehabilitation management. Breast-Rehab integrates a bespoke human action recognition algorithm with a retrieval-augmented generation (RAG) framework. By fusing visual and 3D skeletal data, our model accurately segments exercise videos recorded in uncontrolled home environments, outperforming standard models. These segmented clips, combined with a domain-specific knowledge base, guide a multi-modal large language model to generate clinically relevant assessment reports. This approach significantly reduces computational overhead and mitigates model hallucinations. We implemented the system as a WeChat Mini Program and a nurse-facing dashboard. A preliminary clinical study validated the system's feasibility and user acceptance, with patients achieving an average exercise frequency of 0.59 sessions/day over a two-week period. This work thus presents a complete, validated pipeline for AI-driven, at-home rehabilitation monitoring.

\keywords{Breast Cancer \and Postoperative Rehabilitation \and Human Action Recognition \and Multi-modal Large Language Model \and mHealth.}
\end{abstract}
\section{Introduction}

Up to 82\% of breast cancer survivors may suffer from treatment-related morbidities, especially upper limb dysfunction. Approximately 10\% to 60\% of women report symptoms related to upper limb dysfunction from 2 months to 3 years after surgery \cite{shamleyShoulderMorbidityTreatment2012}, and these symptoms often persist. Clinical experiments have shown that nursing interventions, such as evidence-based care\cite{wangImpactEvidencebasedNursing2024} and continuous personalized care\cite{kongPersonalizedContinuousCare2024}, can effectively improve survivors' quality of life and shoulder mobility. However, the sheer volume of breast cancer survivors means that health institutions lack sufficient human resources to provide such services to all patients. In response, Brennan et al. \cite{brennanPatientExperiencesRehabilitation2020} argued for the urgent need for mHealth systems to help patients adhere to correct postoperative exercises.

In the field of breast cancer limb rehabilitation, some studies have explored using Virtual Reality (VR) to enhance interactivity and guide patients through correct rehabilitation movements. A 2024 meta-analysis by Chen et al.\cite{chenEffectivenessVirtualRealityBased2024} of seven studies in this area concluded that VR-based interventions could improve upper limb function and reduce postoperative pain, but did not show statistically significant differences in lymphedema. The paper noted that the small sample sizes in the field might introduce bias in the conclusions. Furthermore, the high cost of VR equipment currently limits its widespread clinical adoption.

With the development of multi-modal large language models (mm-LLMs), research related to exercise and rehabilitation has begun to incorporate multi-modal information such as vision. The UbiPhysio system, developed by Wang et al. \cite{wangUbiPhysioSupportDaily2024} at Tsinghua University, first collected joint skeletal data for 18 actions such as "sweeping the floor" and "getting out of bed" using multiple sensors. Then it used this skeletal information to classify actions and fed the skeletal data, action classification, and retrieved knowledge base information into a unimodal LLM to provide feedback to the user. Wang et al. \cite{wangRehabGPTMaaSbasedSolution2024} from Zhejiang Lab designed RehabGPT, a system intended to provide action analysis and guidance based on training plans and videos, using image feature extraction, human mesh reconstruction, and text-video matching. However, their paper lacked detailed methodological descriptions and was not tested in a real clinical setting. Sun et al. \cite{sunResearchRehabilitationExercise2024} from Tsinghua University, Shenzhen, designed a Rehabilitation Exercise Guidance System (REGS) to assess the correctness of movements based on temporal skeletal data against standards provided by physical therapists and to offer intervention information. This study included five actions, such as the "glute bridge," and used GPT-generated simulated consultation requests to test the system, comparing its feedback with expert responses.

While these studies have explored the potential of mm-LLMs in the health and fitness domain, no intelligent system specifically for postoperative breast cancer rehabilitation has been developed and applied. For the scenario addressed in this study, postoperative rehabilitation for breast cancer involves numerous exercises, some requiring additional equipment such as resistance bands. Videos uploaded by patients at home often have complex backgrounds and contain a lot of irrelevant information \cite{liQualityAssessmentIntheWild2019}. Directly inputting these videos into an mm-LLM would make it difficult for the model to align action information with the actual video frames, leading to inaccurate assessments and even severe hallucinations \cite{baeMASHVLMMitigatingActionScene2025}, such as evaluating actions that were not performed. This observation is also substantiated in the Figure \ref{fig:errors} of this study. Moreover, the long video lengths impose a significant computational burden, making the system's time and economic costs prohibitive. To address this, both Wang et al. \cite{wangUbiPhysioSupportDaily2024} and Sun et al. \cite{sunResearchRehabilitationExercise2024} employed action recognition algorithms. However, Wang et al.'s research relied on sensor data collection, which is resource-intensive. Sun et al.'s study was conducted in a more controlled environment with a fixed background and less complex actions compared to the at-home management scenario, making direct application of their methods unfeasible.

Furthermore, existing mm-LLMs have not been specifically trained for the breast cancer domain, resulting in generic and superficial evaluations of postoperative rehabilitation exercises that lack sufficient domain knowledge. Therefore, a knowledge-enhancement process is needed to enable the mm-LLM to generate clinically valuable assessment reports. The primary challenge in building this system is to accurately recognize and evaluate the diverse set of upper limb rehabilitation exercises from videos recorded by patients in complex home environments.

To address these challenges, this study developed the Breast-Rehab system, which is based on an action recognition algorithm and an mm-LLM. The system processes postoperative exercise videos uploaded by patients from a WeChat Mini Program, accurately identifying and segmenting key exercise frames. By integrating a breast cancer knowledge base and a retrieval-augmented generation (RAG) algorithm, it employs prompt engineering to drive an mm-LLM to generate a detailed assessment report for each exercise session. This report is displayed on both the patient's interface and a dashboard for healthcare professionals, aiding patients in improving their exercises and facilitating remote monitoring and management.

The main contributions of this study are threefold:
\begin{enumerate}
    \item We designed an action recognition algorithm that fuses image information and 3D skeleton data. On a small-sample dataset of patient postoperative exercise videos, this algorithm outperformed mainstream mm-LLMs in zero-shot and few-shot settings, as well as other existing action classification models.
    \item We conducted a subjective human evaluation of assessment reports generated for 11 real exercise videos and compared the performance with other models. The results showed that the Breast-Rehab system reduces computational overhead, minimizes model hallucinations, and improves report quality.
    \item Building upon the algorithmic achievements, we implemented a patient-facing WeChat Mini Program and a clinician-facing dashboard. A subsequent two-week preliminary clinical study with 15 enrolled patients demonstrated that the system helped users maintain an average exercise frequency of 0.5 sessions per day.
\end{enumerate}

\section{Method}

\subsection{System Workflow}
\begin{figure}
\includegraphics[width=\textwidth]{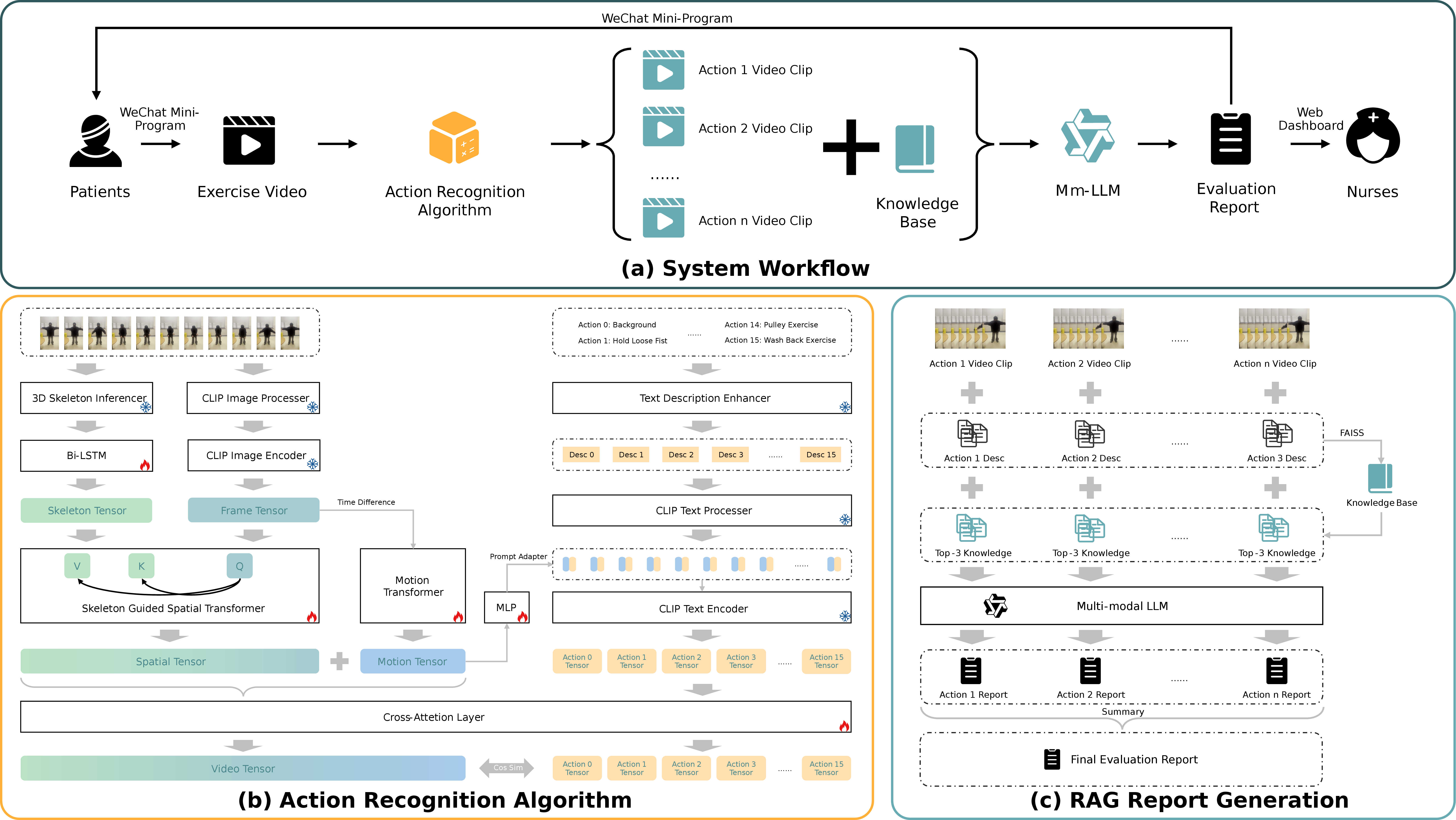}
\caption{Breast-Rehab Method Overview} \label{fig:workflow}
\end{figure}

The workflow of the Breast-Rehab system is illustrated in Figure~\ref{fig:workflow}(a). Patients register and link their postoperative exercise plan through a WeChat Mini Program. They then use their phone's front camera to record and upload their exercise videos to the server. The server invokes an action recognition algorithm to identify and extract action sub-videos, filtering out background noise irrelevant to the exercise. These sub-videos, along with a knowledge base of 1 million chunks developed in our prior Breast-CRAG \cite{chenBreastCRAGBreastCancer2025} study, are fed into an mm-LLM to generate an assessment report. The report is then presented to both the patient and the healthcare provider via the WeChat Mini Program and a web dashboard, respectively.

This study received ethical approval from the Sir Run Run Shaw Hospital, affiliated with Zhejiang University School of Medicine (Approval No. 2025-0529, Acceptance No. 2025-2535-01). All participating patients provided signed informed consent.

Next, we will detail the action recognition and RAG report generation algorithms, followed by the system implementation of Breast-Rehab. Codes are available at https://github.com/Maxin-C/Breast-Rehab.

\subsection{Action Recognition Algorithm}
Our algorithm design was inspired by the work of Wang et al. \cite{wangSeeingFlowingAdapting2023}, incorporating the motion transformer and MCB as core components, and was enhanced with skeleton information to focus the algorithm's attention on the upper body movements, thereby improving action classification accuracy. Our proposed model recognizes human actions by deeply integrating visual, skeletal, and textual modalities through a novel skeleton-guided and motion-prompted architecture. The model takes two primary inputs: a video clip, represented as a sequence of $N_f$ frames $I \in \mathbb{R}^{B \times N_f \times 3 \times H \times W}$, and a corresponding sequence of processed skeleton features $S \in \mathbb{R}^{B \times N_f \times 17}$, where $B$ is the batch size. The skeleton features were meticulously engineered from 3D human pose data extracted using OpenPose \cite{caoOpenPoseRealtimeMultiPerson2021}. For each frame, we selected 13 keypoints corresponding to the upper body. From these keypoints, we computed four biomechanically significant joint angles: the left elbow, right elbow, left shoulder, and right shoulder angles. The angle $\theta_b$ at a joint $b$ formed by adjacent joints $a$ and $c$ was calculated using the dot product between the vectors connecting the joints, based on their 3D coordinates $(\vec{p}_a, \vec{p}_b, \vec{p}_c)$:
\begin{equation}
    \theta_b = \arccos\left(\frac{(\vec{p}_a - \vec{p}_b) \cdot (\vec{p}_c - \vec{p}_b)}{\|\vec{p}_a - \vec{p}_b\| \|\vec{p}_c - \vec{p}_b\| + \epsilon}\right)
\end{equation}
$\epsilon$ is a very small value used to ensure that the denominator is not zero. The final 17-dimensional feature vector for each frame is a concatenation of features derived from the 13 selected keypoints and these four calculated angles.

The model architecture is a dual-stream process generating distinct spatial and motion representations. The spatial stream begins by processing the frame sequence $I$ through a pre-trained CLIP vision encoder, $\Phi_I$, to obtain frame-level visual features, $V_{seq} \in \mathbb{R}^{B \times N_f \times D}$, where $D$ is the embedding dimension. In parallel, the engineered skeleton sequence $S$ is fed into a Skeleton Temporal Encoder, comprising a bidirectional LSTM, to capture its temporal dynamics, yielding skeleton features $K_s \in \mathbb{R}^{B \times N_f \times D}$. These modalities are fused using a Guided Spatial Transformer, which functions as a Transformer decoder, using $V_{seq}$ as the query and $K_s$ as the key and value. This produces a spatially-aware feature sequence $T_s$. Concurrently, the motion stream models temporal changes by applying a Motion Transformer to the pairwise differences of $V_{seq}$, learning a high-level motion representation $T_m \in \mathbb{R}^{B \times N_m \times D}$. The final video representation $V \in \mathbb{R}^{B \times 1 \times D}$ is obtained by summing the mean-pooled features from both streams: $V = \text{AvgPool}(T_s) + \text{AvgPool}(T_m)$.

To achieve a fine-grained, domain-specific alignment between video and text, we utilized class descriptions generated in a novel manner. Instead of generic labels, our text prompts are detailed motion descriptions produced by an mm-LLM. This model was prompted with the official illustrated manual for post-operative upper limb rehabilitation exercises for breast cancer patients, which was originally compiled by nursing staff. This ensures our text modality is rich with clinical context and precise terminology. To make these descriptions video-specific, we inject a dynamic motion prompt $P_m = \text{MotionAdapter}(T_m)$ generated from the video's motion features $T_m$. For each video, this prompt $P_m$ is prepended to the token embeddings of every class description, $[P_m ; E_{text}]$. The concatenated sequence is processed by our custom CLIP text transformer, $\Phi_T^{cus}$, yielding a set of motion-aware text features $T \in \mathbb{R}^{B \times N_c \times D}$ for all $N_c$ classes.

Finally, to further enhance alignment, both the video feature $V$ and text features $T$ are fed into a cross-attention layer. This layer employs two parallel cross-attention layers to refine each modality using information from the other:
\begin{align}
    T' &= \text{LayerNorm}(T + \text{CrossAttention}(Q=T, K=V, V=V)), \label{eq:mcb_text} \\
    V' &= \text{LayerNorm}(V + \text{CrossAttention}(Q=V, K=T, V=T)). \label{eq:mcb_video}
\end{align}
The final classification is performed by computing the cosine similarity between the enhanced video feature $V'$ and text features $T'$. The resulting logits are scaled by a learnable temperature parameter $\tau$ and optimized using a cross-entropy loss.
\begin{equation}
    \text{logits} = \exp(\tau) \cdot \frac{V' \cdot T'}{\|V'\| \|T'\|}
    \label{eq:final_logits}
\end{equation}

\subsection{RAG Report Generation}
The process of RAG report generation is shown in Figure~\ref{fig:workflow}(c). The knowledge base for RAG was derived from the breast cancer knowledge base designed in our previous Breast-CRAG study. This knowledge base contains 1 million chunks, each 100 tokens long, from English and Chinese books and guidelines on breast cancer (including nursing-related literature). These chunks were encoded using a retriever fine-tuned in the Breast-CRAG \cite{chenBreastCRAGBreastCancer2025} study, and a FAISS database was constructed using the chunk embeddings as indices for retrieval.

For each segmented sub-video, the corresponding action description for its action class was identified. The embedding of this text description was then used as a query to search the database, retrieving the top 3 most relevant content chunks as a source for knowledge enhancement. Since the retrieval process in this RAG task is static, we consolidated the retrieved knowledge chunks during the experiment to reduce computational overhead during inference.

Considering the limited video input window of the mm-LLM, we first processed each action's sub-video by feeding it into the mm-LLM along with its corresponding action description and related knowledge to obtain an assessment for that single action. These individual assessments were then integrated into a comprehensive report. The prompt design is detailed in Appendix A.

\subsection{System Implement}
\begin{figure}
\includegraphics[width=\textwidth]{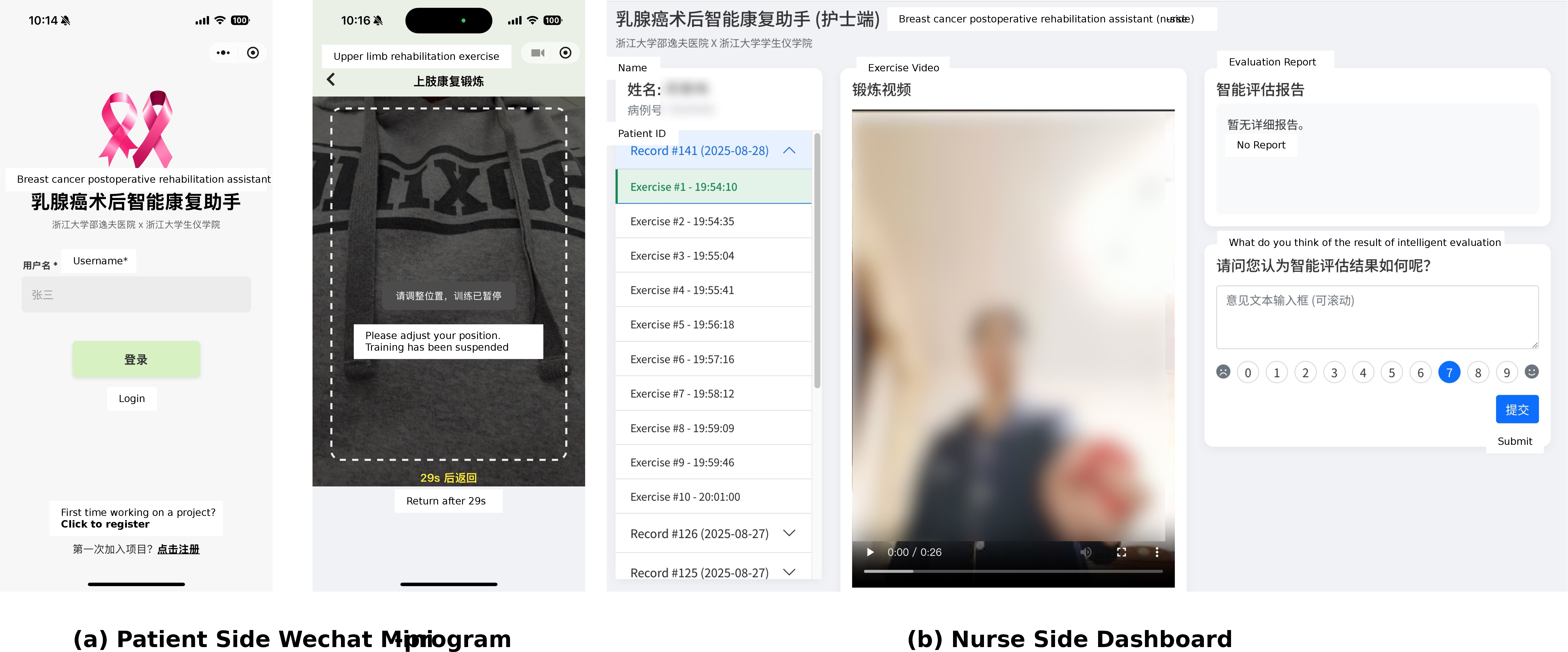}
\caption{Breast-Rehab User Interface} \label{fig:ui}
\end{figure}

The front end of the Breast-Rehab system was developed using HTML, JS, and CSS. The server uses MySQL as the database and Flask as the backend. The system comprises a patient-side WeChat Mini Program and a nurse-side dashboard (as shown in Figure~\ref{fig:ui}).

After registration, users can record their exercise sessions on the exercise page by activating the front camera. During recording, the position of the user's upper body is monitored at a rate of 1 Hz. If the skeleton nodes of the left and right shoulders move out of the camera's field of view, the recording is paused, and the patient is prompted to reposition themselves within the frame (Figure~\ref{fig:ui}(a)). When logging into the system, users can choose whether to subscribe to an exercise reminder at 8 a.m. the next day, helping patients manage their personal exercise progress. The mini-program includes exercise tutorial videos filmed by the nursing team of Sir Run Run Shaw Hospital, allowing patients to learn exercise movements at any time. Figure~\ref{fig:ui}(b) shows the interaction page for a specific patient on the nurse's dashboard, accessible from the patient list. This page displays the user's basic information, exercise records, segmented exercise videos, and the intelligent report. Nurses can evaluate and provide feedback on the generated reports to help with future system optimization.

\section{Experiment}

\subsection{Action Recognition Algorithm}
Based on the definitions and requirements for postoperative upper limb functional exercises for breast cancer patients from the Sir Run Run Shaw Hospital's manual, we included 15 different rehabilitation exercises. Before clinical deployment, we collected 14 portrait-mode videos (30 fps) shot with smartphones in real-world scenarios, where the exerciser's upper body was visible throughout. Using these videos, we manually annotated the action classes on the timeline with the open-source software labelU, including 15 action labels and one "no action" label.

\begin{figure}
\includegraphics[width=\textwidth]{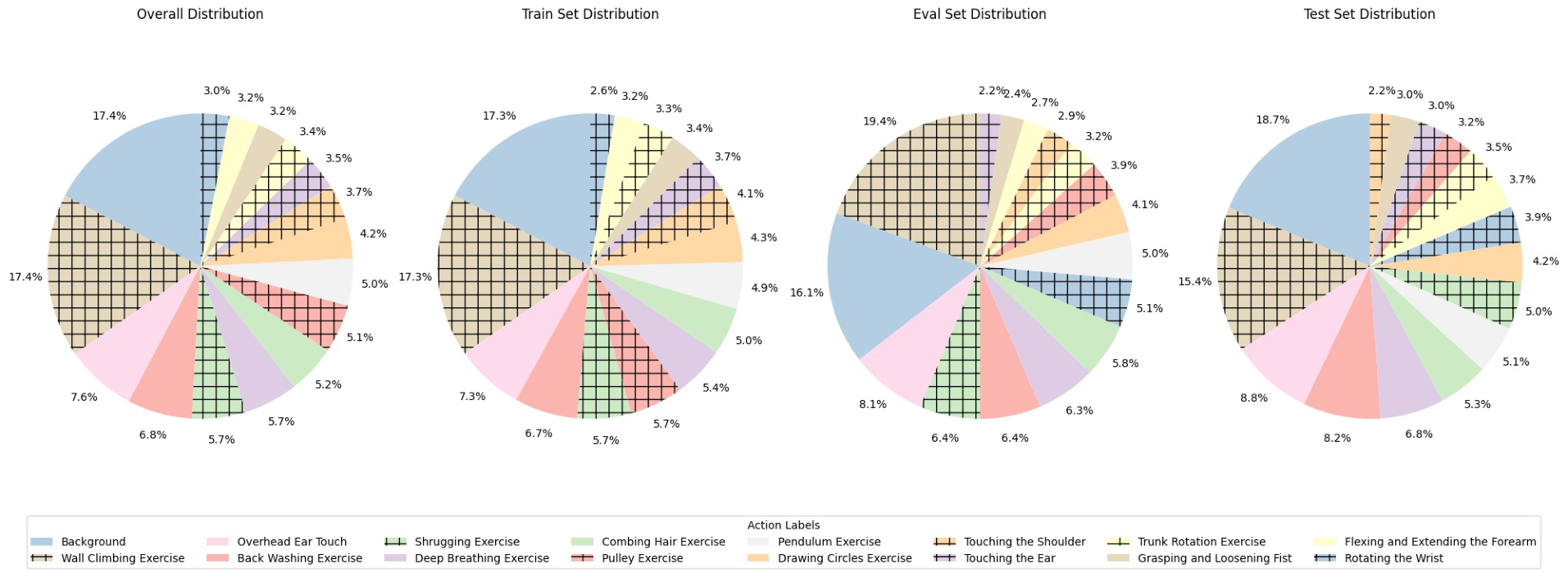}
\caption{Action Label Distribution in the Action Recognition Dataset.} \label{fig:distribution}
\end{figure}

Based on the annotations, we constructed a dataset by extracting video frames using a sliding window approach. The rule was to sample one frame every 6 frames, with a window size of 60 frames and a stride of 21 frames. This resulted in a dataset of 10-frame sequences, where each sequence had a label for each frame and an overall label. If the labels within a 10-frame sequence were inconsistent, the most frequent label was chosen as the overall label. In case of a tie, the action label was prioritized over the "no action" label. The first 10 videos were used as the training set, and the remaining 4 videos were shuffled and split equally into validation and test sets. There was no data overlap between the training and test sets. Ultimately, this process yielded 5,445 training samples and 902 samples each for the validation and test sets. The label distribution of the dataset is shown in Figure~\ref{fig:distribution}.

On the test set, we compared our model against several baselines, including action recognition models from pre-trained mm-LLMs and an established action recognition model. The mm-LLMs included were Qwen-VL-Max (Qwen-VL) \cite{yangQwen3TechnicalReport2025} and GPT-4o-2024-08-06 (GPT-4o) \cite{openaiGPT4TechnicalReport2024}, evaluated in both zero-shot and few-shot settings (prompt designs in Appendix B). The established action recognition model was the one proposed by Wang et al. \cite{wangSeeingFlowingAdapting2023} (abbreviated as CLIP-AR), which we reference. For a fair comparison, this model was trained on our training set. Evaluation metrics included F1-score (weighted), top-1 accuracy, and top-3 accuracy. (Top-3 accuracy was not considered for mm-LLMs as they output direct results).

To validate the value of our algorithm design without repeating the evaluation of the referenced design, we conducted an ablation study. We sequentially removed the skeleton-related components: skeleton information, the Skeleton Temporal Encoder, and the Guided Spatial Transformer. Without skeleton information, our model is identical to CLIP-AR, so we could reuse its results. When the Skeleton Temporal Encoder was removed, an MLP (with a hidden layer dimension matching the encoder) was used to map the 17-dimensional skeleton data to $K_s$. When the Guided Spatial Transformer was removed, an MLP (with a hidden layer dimension matching the transformer) was used to map the concatenated vector of $V_{seq}$ and $K_s$ to a vector with the same dimension as $T_s$.

\subsection{RAG Report Generation}
The dataset for report evaluation consisted of 11 real exercise videos collected from users during clinical application. The objective was to compare the quality of reports generated by an open-source action understanding model, standard mm-LLMs, and mm-LLMs enhanced by the Breast-Rehab system using real-world videos. The open-source action understanding model included was motionLLM \cite{chenMotionLLMUnderstandingHuman2024}, a model released in 2024 capable of understanding actions from video input. The standard mm-LLMs were Qwen-VL and GPT-4o, as in the previous experiment. The Breast-Rehab enhanced models were the same mm-LLMs.

Given that exercise videos are often around 10 minutes long, feeding the entire video into a model at once would overload the input. Therefore, we divided the videos into 45-second chunks, sampling one frame per second, and fed them into the models batch by batch. The results were then aggregated by a large language model (for motionLLM, we used Qwen-VL for aggregation). Similarly, although Breast-Rehab segments the videos into action-specific sub-videos, these can still be too long. Thus, longer sub-videos also followed the batch processing approach before their reports were aggregated. The prompts used for processing sub-videos and integrating their reports are shown in Appendix A.

Due to the lack of a benchmark, the reports were evaluated manually. We asked nurses from the breast surgery department at Sir Run Run Shaw Hospital to first watch the videos and then evaluate the content generated by the five models. The evaluation was single-blinded; the evaluators were unaware of which model produced which content. A 10-point Likert scale was used for quantitative assessment (see Table~\ref{tab:likert}). A higher score indicates greater agreement with the statement, signifying better model performance on that dimension. Data was collected via Tencent Questionnaire, and all user faces in the videos were blurred.

\begin{table}
\caption{Likert Scale Design for Human Evaluation of Reports.}\label{tab:likert}
\centering
\begin{tabularx}{\textwidth}{p{0.2\textwidth}X} 
\toprule
\textbf{Dimension} & \textbf{Content} \\
\midrule
Accuracy & The model accurately identified and described the errors that occurred in the exercise videos. \\
Completeness & Compared to the critical issues you observe in the video, the model does not appear to miss significant points. \\
Practicability & The corrective suggestions given by the model are specific and practical. \\
Safety & There is no potential risk of harm in the assessment recommendations given by the model. \\
Language quality & The assessment report is fluent and easy to understand. \\
\bottomrule
\end{tabularx}
\end{table}

Based on the evaluation results, we first performed descriptive statistics on the mean and variance of the scores for each model across the five dimensions. Then, after conducting a Shapiro-Wilk test for normality, we used either an independent samples t-test or a Mann-Whitney U test to compare whether there was a statistically significant difference in the scores of the same mm-LLM with and without Breast-Rehab enhancement.

\subsection{System Implement}
We promoted the system and recruited postoperative breast cancer patients for a two-week study from August 23, 2025, to September 6, 2025, by distributing flyers. To verify patients' compliance with upper limb rehabilitation exercises for breast cancer under the Breast-Rehab system, this study objectively measures this indicator based on the average number of exercise sessions and the average exercise frequency. The average number of exercise sessions is calculated by dividing the total number of exercise sessions recorded by all enrolled patients using the system during the experimental period by the total number of enrolled patients. The exercise frequency is calculated by dividing the number of exercise sessions of a single patient by the number of days they were enrolled. The average exercise frequency is obtained by dividing the total exercise frequency by the total number of enrolled patients. For subjective evaluation, this study collected feedback from nurses and patients who used the Breast-Rehab system and summarized the current strengths and weaknesses of the system.

\section{Result}

\subsection{Action Recognition Algorithm}
\begin{figure}
\includegraphics[width=\textwidth]{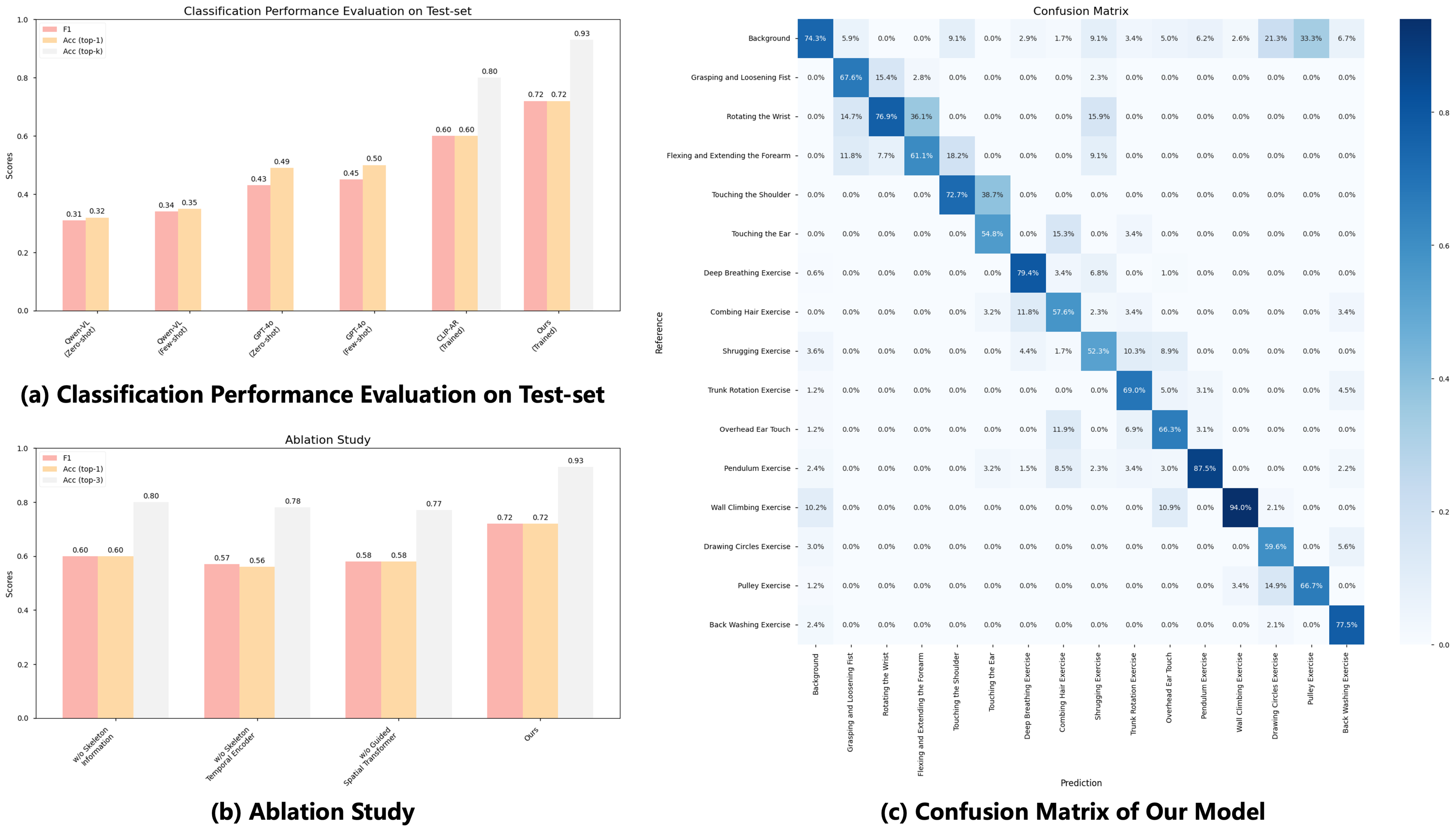}
\caption{Action Recognition Algorithm Evaluation Result} \label{fig:action_rec_results}
\end{figure}

The evaluation results on the test set are shown in Figure~\ref{fig:action_rec_results}(a). The results indicate that Qwen-VL and GPT-4o achieved an accuracy of less than 50\% on this dataset. Although few-shot learning with contextual information improved accuracy by about 10\%, it was still insufficient for practical application. Furthermore, the inference speed of these models was excessively slow for long contexts and multiple video inputs. Qwen-VL (few-shot) averaged 11 seconds per data point (10 frames), while GPT-4o (few-shot) averaged 140 seconds per case. Even considering that GPT-4o's performance might be affected by network issues in China, the inference time of the domestic mm-LLM was still prohibitive. After training on the training set, the CLIP-AR model achieved a better accuracy (0.60) than the mm-LLMs, with a top-3 accuracy of 0.80. By incorporating skeleton information, our model's accuracy improved by about 10\%, reaching 0.72.

The confusion matrix for our model (showing precision percentages) is depicted in Figure~\ref{fig:action_rec_results}(c). The model achieved a precision of over 50\% for all classes. The three actions with the lowest precision were "touching the ear" (0.548), "combing hair exercise" (0.576), and "shrugging exercise" (0.523). "Touching the ear" was most frequently misclassified as "touching the shoulder". This is partly because these two exercises can be performed interchangeably according to the manual, leading to frequent switching between them in the dataset, which biased the model towards predicting "touching the shoulder". Additionally, the only difference between these two actions is whether the affected limb touches the ipsilateral ear or the contralateral shoulder, with other body parts remaining consistent. This distinction is even less clear when the patient alternates between the two, making it difficult for the model to differentiate. "Combing hair exercise" was misclassified as "touching the ear" and "overhead ear touch". The arm movement in "combing hair exercise" is similar to the other two, with the hand just positioned slightly higher (touching the top of the head instead of the ear). The model needs to precisely distinguish the hand's position relative to the head to separate these actions, hence the lower precision. The challenge with "shrugging exercise" lies in its subtle range of motion. During shrugging, only the shoulder joint moves slightly. When recorded from the front, both skeleton and image data show only minor vertical movement of the shoulder, providing very weak features and causing the model to misclassify it as other actions.

\subsection{Report Evaluation}
For the human evaluation, five nurses assessed the outputs of five models on forms created from 11 sets of videos. The evaluation results for the reports from the five models are shown in Table~\ref{tab:human_eval}. The results indicate that motionLLM performed the worst in this experiment. From a mean score perspective, Breast-Rehab improved the performance of both Qwen-VL and GPT-4o across all five dimensions. Furthermore, Qwen-VL performed better than GPT-4o in both with and without Breast-Rehab scenarios and achieved the highest scores on four of the five metrics.

\begin{table}
\caption{Human Evaluation Result (The values in the table are the mean scores, with variances in parentheses. Bold entries are the optimal scores).}\label{tab:human_eval}
\centering
\begin{tabularx}{\textwidth}{l YYYYY}
\toprule
 & \textbf{Accuracy} & \textbf{Complete-\newline ness} & \textbf{Practicability} & \textbf{Safety} & \textbf{Language \newline quality} \\
\midrule
MotionLLM & 5.69 (1.98) & 5.12 (1.95) & 5.73 (1.97) & 5.92 (2.11) & 7.10 (2.01) \\
Qwen-VL & 7.55 (1.47) & 7.49 (1.17) & 7.42 (1.50) & 7.59 (1.37) & 7.66 (1.45) \\
GPT-4o & 7.05 (2.30) & 7.48 (1.67) & 7.30 (2.14) & 7.40 (1.03) & 7.63 (1.09) \\
Qwen-VL (w BR) & \textbf{7.89} (2.25) & \textbf{8.33} (2.25) & \textbf{8.00} (1.78) & 7.89 (3.36) & \textbf{8.00} (0.75) \\
GPT-4o (w BR) & 7.87 (2.45) & 7.93 (2.55) & 7.64 (1.27) & \textbf{7.93} (1.58) & 7.53 (1.98) \\
\bottomrule
\end{tabularx}
\end{table}

A Shapiro-Wilk test was used to assess the normality of the evaluation results for the four models excluding motionLLM. The results showed that all p-values were greater than 0.05, indicating that the data did not follow a normal distribution. Therefore, the Mann-Whitney U test was used to compare the performance difference of the models with and without Breast-Rehab enhancement. The results are shown in Table~\ref{tab:mann_whitney}. The results indicate that with Breast-Rehab enhancement, both models achieved statistically significant improvements in Accuracy, Completeness, Practicability and Safety, but not in Language quality.

\begin{table}
\caption{P-values of Mann-Whitney U test (Bold values indicate statistical differences, $\alpha$=0.05).}\label{tab:mann_whitney}
\centering
\begin{tabularx}{\textwidth}{l YYYYY}
\toprule
 & \textbf{Accuracy} & \textbf{Complete-\newline ness} & \textbf{Practicability} & \textbf{Safety} & \textbf{Language \newline quality} \\
\midrule
Qwen-VL & \textbf{0.0415} & \textbf{0.0028} & 0.5086 & \textbf{0.0496} & 0.5037 \\
GPT-4o & \textbf{0.0008} & 0.2577 & \textbf{0.0241} & 0.1212 & 0.7306 \\
\bottomrule
\end{tabularx}
\end{table}

\begin{figure}
\includegraphics[width=\textwidth]{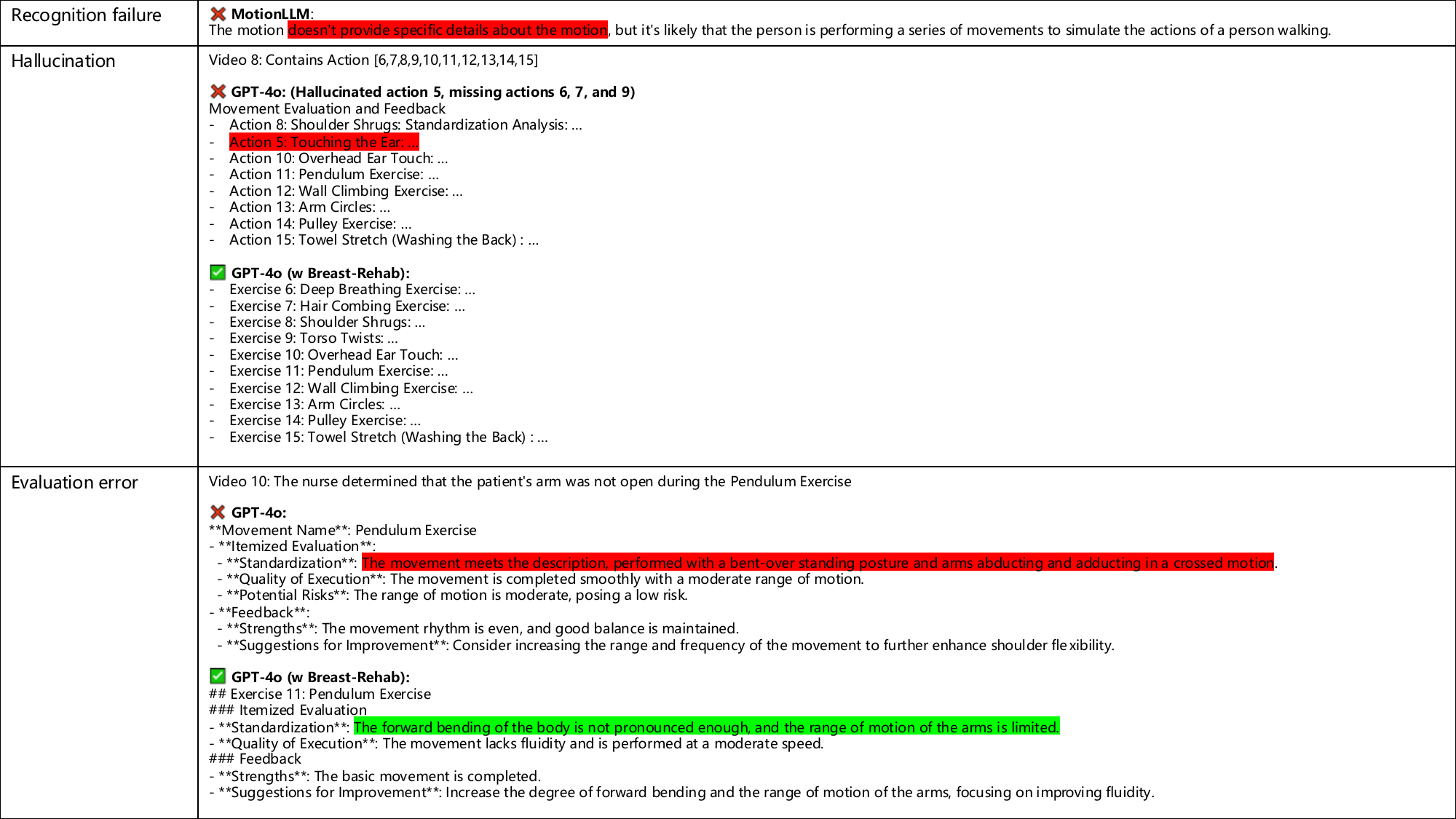}
\caption{Typical Error Examples of mm-LLMs without Breast-Rehab Enhancement.} \label{fig:errors}
\end{figure}

A detailed analysis of the output content revealed several issues with models that did not have Breast-Rehab enhancement, as shown in Figure~\ref{fig:errors}: 1) Recognition Failure: The model struggled to correctly understand and extract information from the video content. This was most evident in motionLLM's analysis. The initial failure in recognition led to a lack of informative content in the final report, resulting in poor quality. 2) Hallucination: For standard mm-LLMs, due to their limited action recognition capabilities, they often misclassified action types during segmented analysis. This led to the hallucination of non-existent actions and their evaluation in the summary report. 3) Evaluation Error: Without the enhancement of additional action descriptions and related knowledge, the mm-LLM might judge an incorrect action as correct, leading to erroneous evaluations in the final report.

\subsection{System Implement}
During the experiment, flyers were distributed to a total of 25 patients, 15 of whom joined the Breast-Rehab program, representing a participation rate of 60\% (15/25). According to the definition of objective indicators in the Experiment section, the total number of exercise sessions recorded by patients during the study was 57, with an average of 3.8 sessions per patient (57/15). Four patients exercised more than 5 times, with the highest number of sessions being 12. The average exercise frequency was 0.59. The highest and lowest exercise frequencies were 1 session/day (3/3) and 0.25 sessions/day (1/4), respectively. This indicates that, on average, patients used the system for upper limb rehabilitation exercises once every two days. The best-case scenario involved daily exercise, while the worst-case scenario involved exercising only once over four days.

Qualitative feedback on the Breast-Rehab application was gathered through discussions with two enrolled patients and three nurses. From the patients' perspective, the greatest benefits of the system were its ability to provide exercise reminders and guidance, as well as the convenience of starting exercises at any time once they had developed a habit of using it. However, for patients who had just undergone surgery, the limited range of postoperative exercises and the lack of adequate space for exercising during hospitalization made it difficult for them to persist with using Breast-Rehab. The nurses noted that remotely monitoring patients through analytical reports allowed for a more quantitative understanding of their rehabilitation progress. This data not only facilitated better patient management but also held potential for further clinical research. Additionally, the nurses believed that providing real-time exercise guidance and feedback tailored to patients' treatment progress would further enhance user engagement.

\section{Discussion}
\subsection{Action Recognition Algorithm}
The algorithm evaluation results show that directly inputting video and prompts into current large language models suffers from two main issues: insufficient accuracy and high time and economic costs. The low accuracy may be due to the difficulty of fully describing an action with prompts, making it hard for the model to correspond. The few-shot approach improved accuracy to some extent because video-based examples better described the action categories. Furthermore, there is a gap between the embedding spaces of text prompts and video or image sequences. For the specific research scenario in this study, large models have been trained on very little relevant data, making image classification based solely on model performance difficult. By training on a small number of labeled samples, even with completely different video recording environments between the training and test sets, the model can achieve good classification performance in a specific scenario, which is intuitive.

The ablation study results indicate that adding skeleton information to the reference algorithm allows the model to focus more on the specific actions, reducing interference from environmental information and thus significantly improving accuracy. Additionally, the ablation study showed that simply mapping skeleton information with an MLP is ineffective and even detrimental to the model. Directly transforming the vector dimensions instead of using a Skeleton Temporal Encoder prevents the expression of temporal information in the skeleton data. Similarly, simply concatenating vectors instead of using a Guided Spatial Transformer fuses skeleton information as noise into the image information, not only failing to increase focus on the person but also interfering with the input vector information.

The confusion matrix results show that the model's performance significantly drops for classification tasks with small motion amplitudes or similar actions. Such tasks require the model to filter out key subtle perturbations from environmental noise for accurate recognition. Therefore, adding extra feature enhancement information, such as optical flow to focus on changes in person's movements, or increasing features of the relative positions of skeleton nodes, could potentially improve model performance, but would also increase the model's overhead. Moreover, the training set included in this experiment was relatively small for the number of action categories. Increasing model complexity on a small sample size will lead to overfitting. Therefore, future research could consider continuously collecting and annotating data during experiments to form a larger benchmark to improve model performance.

\subsection{RAG Report Generation}
The human evaluation of the reports shows that action understanding models like motionLLM have limited ability to understand complex videos and instructions in application scenarios beyond their benchmarks. Among the five dimensions, only language quality was close to that of other models. However, this report was generated by Qwen-VL integrating motionLLM's analysis of video clips, so the higher score in this aspect is more dependent on the summarizing model's polishing of the report. Among the typical errors, motionLLM's inability to recognize video content was particularly prominent, making it unsuitable for direct use as a local report processor.

By comparing the performance of Qwen-VL and GPT-4o with and without Breast-Rehab, it is evident that the system proposed in this study can effectively improve the accuracy of the reports generated by the model, but it has little impact on the language quality of the report itself, as language quality is more related to the language generation capabilities of the large model itself. With clear action classification, mm-LLMs are much less likely to miss or hallucinate action content, although this is limited by the accuracy of the action recognition model. Furthermore, by filtering out irrelevant background information, the model's inference overhead is also greatly reduced.

\subsection{System Implement}
The preliminary clinical study demonstrates that it helps patients adhere to exercise at an average frequency of approximately once every two days, with the maximum number of exercise sessions reaching 12. This indicates that the system operates stably and effectively guides patients in consistently performing rehabilitation exercises. However, based on patient feedback, those who have just undergone surgery often struggle with compliance due to inadequate hospital environments, highlighting the need for enhanced in-hospital education and improved system interaction. These improvements would help patients better understand the value of Breast-Rehab and derive positive feedback from using the system. The nurses' suggestions regarding real-time interaction can further enhance the system’s interactive performance, though generating personalized interactive content and achieving high-speed inference capabilities—key to enabling real-time interaction—pose significant challenges for the next phase of Breast-Rehab’s development.

However, it must be acknowledged that this experiment lacks rigorous control groups to compare changes in various indicators before and after using Breast-Rehab. Currently, we are conducting a randomized controlled trial (RCT) involving more than 50 participants, monitoring indicators such as joint range of motion over a six-week period. The study will ultimately validate the system's usability. The experimental results will be made publicly available, and related papers will be published in due course.


\section{Conclusion}
This study, in the field of upper limb rehabilitation exercises after breast cancer surgery, designed a motion recognition algorithm incorporating skeletal information to improve accuracy, and established a RAG-based report generation framework that integrates a knowledge base with motion recognition data. This approach reduces hallucinations in rehabilitation exercise assessment reports generated by mm-LLMs while also decreasing time and economic costs. Based on the aforementioned algorithms, a clinically applicable system was developed and deployed in a hospital for a two-week preliminary clinical study. The system helped patients maintain an average upper limb rehabilitation exercise frequency of once every two days, demonstrating its usability. This work paves the way for further clinical application and improvement of postoperative rehabilitation outcomes for patients.

%
%
%
\bibliographystyle{splncs04}
\bibliography{references}

\appendix
\section{Breast-Rehab RAG Report Generation Prompt Design}
\small
\begin{verbatim}
Single action prompt:
# Role
You are a professional post-operative breast cancer rehabilitation trainer.
# Task
Please evaluate a video of a [specific individual action]. Your evaluation
needs to be highly focused and specific.
# Evaluation Subject
* Action Name: Action {action_id}
* Standard Action Description: {action_desc}
# Evaluation Instructions
Based on the standard description above and the series of video frames
provided next, conduct a detailed evaluation of the patient's action
from the following three dimensions:
1. Action Standardization: Does the posture, range of motion, and
   trajectory comply with the standard?
2. Completion Quality: Is the action complete and smooth? Is the speed
   appropriate? Is there any compensation?
3. Feedback and Suggestions: Clearly point out what was done well (strengths)
   and areas needing improvement (suggestions for improvement).
Please present your analysis in clear bullet points.

Chunk evaluation prompt:
# Role
You are a professional rehabilitation trainer analyzing a
[segment of a long video].
# Task
Please analyze only the patient's performance in the current video segment.
Your analysis will be used for subsequent summarization.
# Evaluation Subject
* Action Name: Action {action_id}
* Standard Action Description: {action_desc}
# Segment Analysis Instructions
Briefly describe the observed standardization of the patient's movement,
completion quality, strengths, and areas for improvement in this segment.

Action synthesis prompt:
# Role
You are a rehabilitation trainer tasked with synthesizing multiple video
segment analyses of a [single action].
# Task
The following are analyses of consecutive segments from a [long video].
Please integrate them into a coherent and complete evaluation report
for this [one action].
# Evaluation Subject
* Action Name: Action {action_id}
* Standard Action Description: {action_desc}
# Synthesis Instructions
Synthesize the observations from all segments to summarize the overall
performance of this action, including standardization, quality, strengths,
and suggestions for improvement.
# Below are the independent analyses of all segments:
{chunk_evaluations}

Final synthesis prompt:
# Role
You are a senior post-operative breast cancer rehabilitation trainer,
tasked with writing a summary report for a patient's rehabilitation training.
# Task
You have received independent evaluation reports for multiple rehabilitation
actions completed by the patient. Please synthesize these reports into a
comprehensive and coherent final summary.
# Summary Instructions
1. Review all independent evaluations to understand the patient's overall
   performance.
2. In the "Overall Overview" section, summarize the patient's common
   strengths and the most critical recurring issues.
3. In the "Detailed Breakdown of Individual Actions" section, organize the
   independent evaluation results for each action.
4. Ensure the report is professional, positive, and encouraging in tone.
# Below are all the independent evaluation reports:
{all_evaluations}
# Please generate the final summary report based on the above information.
\end{verbatim}

\section{Action Recognition Zero-shot and Few-shot Prompt Design}
\small
\begin{verbatim}
Zero-shot prompt:
You are a helpful expert in action recognition. Your task is to identify
the action being performed by the person in the video. Please select the
most appropriate description from the numbered list provided below.

Action options:
{class_list_str}

Now, please identify the action in this new video.

Your response must only contain the action number. For example: 3

Few-shot prompt:
You are a helpful expert in action recognition. Your task is to identify
the action being performed by the person in the video. Please select the
most appropriate description from the numbered list provided below.

Action options:
{class_list_str}

Here are some examples:
{action_1 video} Action: 1
{action_2 video} Action: 2
...
{action_n video} Action: n

Now, please identify the action in this new video.

Your response must only contain the action number. For example: 3
\end{verbatim}

\end{document}